\documentclass[english,aps,prd,a4paper,groupedaddress,preprintnumbers,floatfix,nofootinbib]{revtex4-1}
\usepackage[english]{babel}
\usepackage{amsmath}
\usepackage{graphicx}
\usepackage{color}
\usepackage{mathrsfs}   
\usepackage{amssymb}
\usepackage{hyperref}
\usepackage{dcolumn}
\usepackage{bm}
\usepackage{graphicx}

\newcommand {\be} {\begin{equation}}
\newcommand {\ee} {\end{equation}}

\def\vev#1{\left\langle #1\right\rangle}

\usepackage{float}

\def\lnv{lepton number violation }
\def\vev#1{\left\langle #1\right\rangle}

\def\e6{$\mathrm{E(6)}$ }
\def\10{$\mathrm{SO(10)}$ }
\def\21{$\mathrm{SU(2)_L \otimes U(1)_Y}$ }
\def\31{$\mathrm{SU(3)_c \otimes U(1)_Q}$ }
\def\SM{$\mathrm{SU(3)_c \otimes SU(2)_L \otimes U(1)_Y}$ }
\newcommand{\sm}{{Standard Model }}

\def\3211{$\mathrm{SU(3) \otimes SU(2)_L \otimes U(1)_R \otimes U(1)_{B-L}}$ }
\def\321{$\mathrm{SU(3) \otimes SU(2) \otimes U(1)}$ }
\def\422{$\mathrm{SU(4) \otimes SU(2) \otimes SU(2)_R}$ }

\newcommand{\AddrAHEP}{%
  AHEP Group, Institut de F\'{i}sica Corpuscular --
  C.S.I.C./Universitat de Val\`{e}ncia, Parc Cient\'ific de Paterna.\\
 C/ Catedr\'atico Jos\'e Beltr\'an, 2 E-46980 Paterna (Valencia) - SPAIN}

\begin{document}

\title{Vacuum stability with spontaneous violation of lepton number}

\author{
Cesar Bonilla,\footnote{Electronic address: cesar.bonilla@ific.uv.es}
Renato M. Fonseca \footnote{Electronic address: renato.fonseca@ific.uv.es}
and\ \ Jos\'e W. F. Valle\footnote{Electronic address: valle@ific.uv.es}}

\affiliation{
\AddrAHEP}
\date{\today}

\begin{abstract} 
  The vacuum of the Standard Model is known to be unstable for the
  measured values of the top and Higgs masses. Here we show how vacuum
  stability can be achieved naturally if lepton number is violated
  spontaneously at the TeV scale. More precise Higgs measurements in
  the next LHC run should provide a crucial test of our symmetry
  breaking scenario. In addition, these schemes typically lead to
  enhanced rates for processes involving lepton flavour violation.
\end{abstract}
\maketitle

\section{Introduction}

The vacuum of the Standard Model (SM) scalar potential is unstable
since at high energies the Higgs effective quartic coupling is driven
to negative values by the renormalization group flow
\cite{Alekhin:2012py,Buttazzo:2013uya}. Nevertheless, the SM cannot be
a complete theory of Nature for various reasons, one of which is that
neutrinos need to be massive in order to account for neutrino
oscillation results~\cite{Forero:2014bxa}.\footnote{Planck scale
  physics could also play a role~\cite{Branchina:2013jra}.}

\begin{figure}[tbph]
\begin{centering}
\includegraphics[scale=.8]{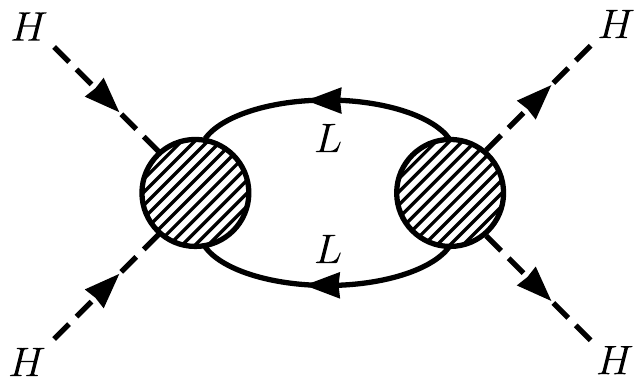}
\par\end{centering}
\protect\caption{\label{fig:1}Contribution of Weinberg's effective
  operator to the Higgs quartic interaction. }
\end{figure}

With only the SM fields, neutrino masses can arise in a
model-independent way from a dimension 5 effective operator $\kappa
LLHH$ which gives rise to a $\kappa \vev{H}^{2}$ neutrino mass after
electroweak symmetry breaking \cite{Weinberg:1979sa}. This same
operator unavoidably provides a correction to the Higgs self-coupling
$\lambda$ below the scale of the mechanism of neutrino mass generation
through the diagram in Fig. \ref{fig:1}. Although tiny\footnote{The
  contribution to $\lambda$ is suppressed by a factor
  $\left(m_{\nu}/\left\langle H\right\rangle
  \right)^{2}/\left(4\pi\right)^{2}$.}
and negative, it suggests that the mechanism responsible for
generating neutrino masses and \lnv is potentially relevant for the
Higgs stability problem. The quantitative effect of neutrino masses on
the stability of the scalar potential will, however, be dependent on
the ultra-violet completion of the model.

After the historic Higgs boson discovery at CERN and the confirmation
of the Brout-Englert-Higgs mechanism, it is natural to imagine that
all symmetries in Nature are broken spontaneously by the vacuum
expectation values of scalar fields. The charge neutrality of
neutrinos suggests them to be Majorana
fermions~\cite{Schechter:1980gr}, and that the smallness of their mass
is due to the feeble breaking of lepton number symmetry. Hence we need
generalized electroweak breaking sectors leading to the double
breaking of electroweak and lepton number symmetries.

In this letter we examine the vacuum stability issue within the
simplest of such extended scenarios~\footnote{Extended Higgs scenarios
  without connection to neutrino mass generation schemes have been
  extensively discussed, see for example, Ref.~\cite{Costa:2014qga}
  and references therein.}, showing how one can naturally obtain a
fully consistent behavior of the scalar potential at all scales for
lepton number broken spontaneously at the TeV scale.  Note that within
the simplest \SM gauge structure lepton number is a global symmetry
whose spontaneous breaking implies the existence of a physical
Goldstone boson, generically called majoron and denoted $J$, which
must be a gauge singlet~\cite{chikashige:1981ui,Schechter:1981cv} in
order to comply with LEP restrictions~\cite{Agashe:2014kda}. Its
existence brings in new invisible Higgs boson
decays~\cite{Joshipura:1992hp}
$$H\to JJ , $$
leading to potentially sizable rates for missing momentum signals at
accelerators~\cite{deCampos:1997mf,Abdallah:2004wy,abdallah:2003ry}
including the current LHC~\cite{Bonilla:2015uwa}. Given the agreement
of the ATLAS and CMS results with the SM scenario, one can place
limits on the presence of such invisible Higgs decay channels.
Current LHC data on Higgs boson physics still leaves room to be
explored at the next
run.

\begin{figure}[tbph]
	\begin{centering}
		\includegraphics[scale=.8]{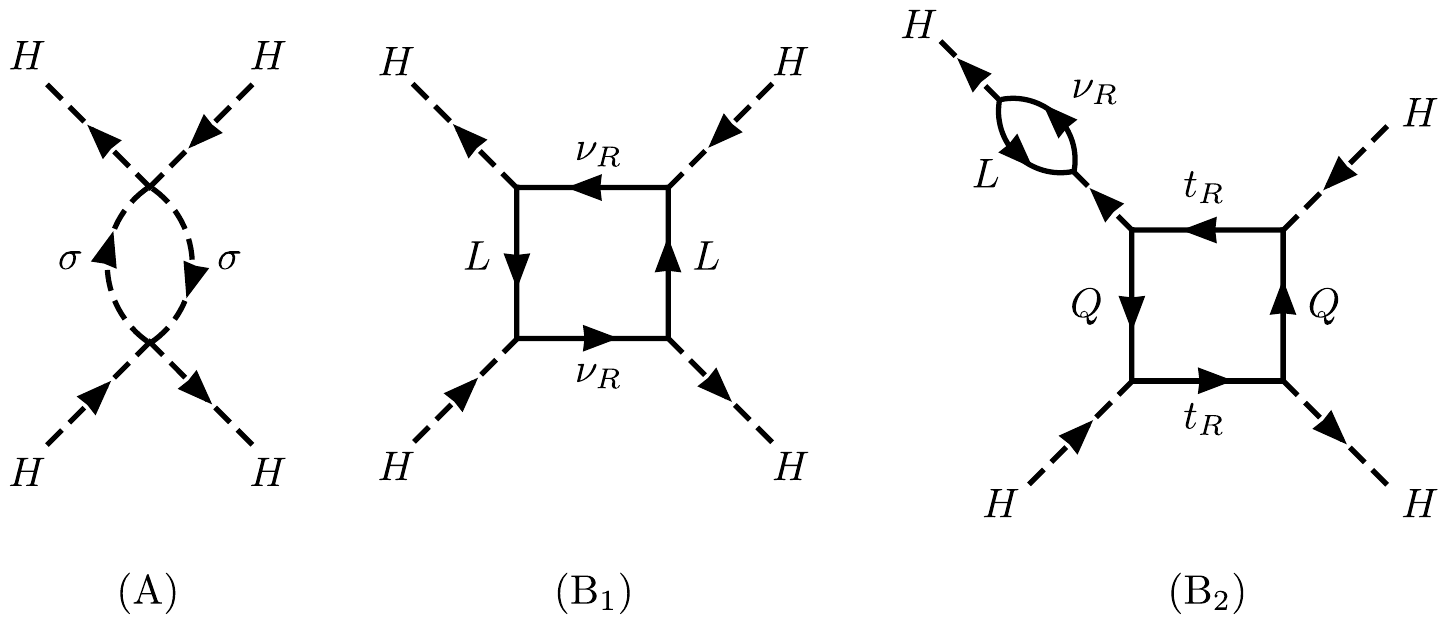}
                  \par\end{centering}
                \protect\caption{\label{fig:2} In models with a
                  complex singlet scalar $\sigma$, such as majoron
                  type-I seesaw schemes, the positive contribution to
                  the RGE of the Higgs quartic coupling (diagram A) is
                  accompanied by the destabilizing effect of
                  right-handed neutrinos through the 1-loop diagram
                  $\textrm{B}_{1}$ and also through the two-loop
                  diagram $\textrm{B}_{2}$.}
\end{figure}

Absolute stability of the scalar potential is attainable as a result
of the presence of the Majoron, which is part of a complex scalar
singlet. Indeed, it is well known that generically the quartic
coupling which controls the mixing between a scalar singlet and the
Higgs doublet contributes positively to the value of the Higgs quartic
coupling (which we shall call $\lambda_{2}$) at high energies
\cite{Casas:1999cd,EliasMiro:2011aa,EliasMiro:2012ay,Basso:2010jm,Gonderinger:2009jp,Lebedev:2012zw,Falkowski:2015iwa,Ballesteros:2015iua,
  Rose:2015fua} --- see diagram A in figure \ref{fig:2}. On the other
hand, new fermions coupling to the Higgs field $H$, such as
right-handed neutrinos
\cite{EliasMiro:2012ay,Salvio:2015cja,Casas:1999cd}, tend to
destabilize $\lambda_{2}$ not only through the 1-loop effect depicted
in diagram $\textrm{B}_{1}$ of figure \ref{fig:2}, but also in what is
effectively a two-loop effect (diagram $\textrm{B}_{2}$): through
their Yukawa interaction with $H$, the new fermions soften the fall of
the top Yukawa coupling at higher energies, which in turn contributes
negatively to $\lambda_{2}$~\footnote{Even though it does not happen
  in our case, one should keep in mind that fermions alone could in
  principle stabilize the Higgs potential by increasing the value of
  the gauge couplings at higher energies, which in turn have a
  positive effect on the Higgs quartic coupling.}.
The model we consider below is a low--scale version of the standard
type I majoron seesaw mechanism, such as the inverse seesaw
type~\cite{mohapatra:1986bd,gonzalezgarcia:1989rw}.
We stress however that, even though our renormalization group
equations (RGEs) are the same as those characterizing standard case,
the values of the Dirac--type neutrino Yukawa couplings are typically
much higher in our inverse seeaw scenario.

\section{Electroweak breaking with spontaneous \lnv}

The simplest scalar sector capable of driving the double breaking of
electroweak and lepton number symmetry consists of the SM doublet $H$
plus a complex singlet $\sigma$, leading to the following Higgs
potential~\cite{Joshipura:1992hp}
\begin{eqnarray}
\label{eq:pot}
& & V\left(\sigma,H\right)=\mu_{1}^{2}\left|\sigma\right|^{2}+
\mu_{2}^{2}H^{\dagger}H+\lambda_{1}\left|\sigma\right|^{4}\notag\\
& &\ \ +\lambda_{2}\left(H^{\dagger}H\right)^{2}+
\lambda_{12}\left(H^{\dagger}H\right)\left|\sigma\right|^{2}\,.
\end{eqnarray}
In addition to the \SM gauge invariance, $V\left(\sigma,H\right)$ has
a global U(1) symmetry which will be associated to lepton number
within specific model realizations.  The potential is bounded from
below provided that $\lambda_{1}$, $\lambda_{2}$ and
$\lambda_{12}+2\sqrt{\lambda_{1}\lambda_{2}}$ are positive; these are
less constraining conditions than those required for the existence of
a consistent electroweak and lepton number breaking vacuum where both
$H$ and $\sigma$ adquire non-zero vacuum expectation values ($\equiv
\frac{v_H}{\sqrt{2}}$ and $\frac{v_\sigma}{\sqrt{2}}$).  For that to
happen, $\lambda_{1}$, $\lambda_{2}$ and
$4\lambda_{1}\lambda_{2}-\lambda_{12}^{2}$ need to be all
positive~\footnote{However, this last condition need not hold for
  arbitrarily large energy scales. Indeed, it is enough to consider
  $4\lambda_{1}\lambda_{2}-\lambda_{12}^{2}>0$ for energies up to
  $\Lambda\approx\textrm{Max}\left(\sqrt{2\frac{\left|\mu_{1}^{2}\right|}{\lambda_{12}}},\sqrt{\frac{\left|\mu_{2}^{2}\right|}{\lambda_{2}}}\right)$
  --- see \cite{EliasMiro:2012ay,Ballesteros:2015iua} for details.}.
Three of the degrees of freedom in $H$ are absorbed by the massive
electroweak gauge bosons, as usual. On the other hand, the imaginary
part of $\sigma$ becomes the Nambu-Goldstone boson associated to the
breaking of the global lepton number symmetry, therefore it remains
massless. As for the real oscillating parts of $H^{0}$ and $\sigma$,
these lead to two CP-even mass eigenstates $H_{1}$ and $H_{2}$, with a
mixing angle $\alpha$ which can be constrained from LHC
data~\cite{Aad:2014iia,Chatrchyan:2014tja,CMS-PAS-HIG-14-038,Bonilla:2015uwa}.
We take the lighter state $H_{1}$ to be the 125 GeV Higgs particle
recently discovered by the CMS and ATLAS collaborations.

Using the renormalization group equations (given in the appendix) we
evolved the three quartic couplings of the model imposing the vacuum
stability conditions mentioned previously. Given that such equations
rely on perturbation theory, the calculations were taken to be
trustable only in those cases where the running couplings do not
exceed $\sqrt{4\pi}$. \footnote{Since all the new particles present in
  the low-scale seesaw model under consideration have yet to be
  observed, leading order calculations suffice.  For our plots we have
  used the values $\alpha_S\approx0.1185$ and $y_t\approx0.96$ at the
  $m_Z$ scale --- more precise values with higher order corrections
  can be found in \cite{Degrassi:2012ry}. Small changes to these input
  values (for example a change of 0.03 in the top Yukawa $y_t$) do not
  affect substancially our plots.}

 \begin{figure}[H]
\vspace{-10pt}
  \centering
  {\includegraphics[width=0.48\textwidth]{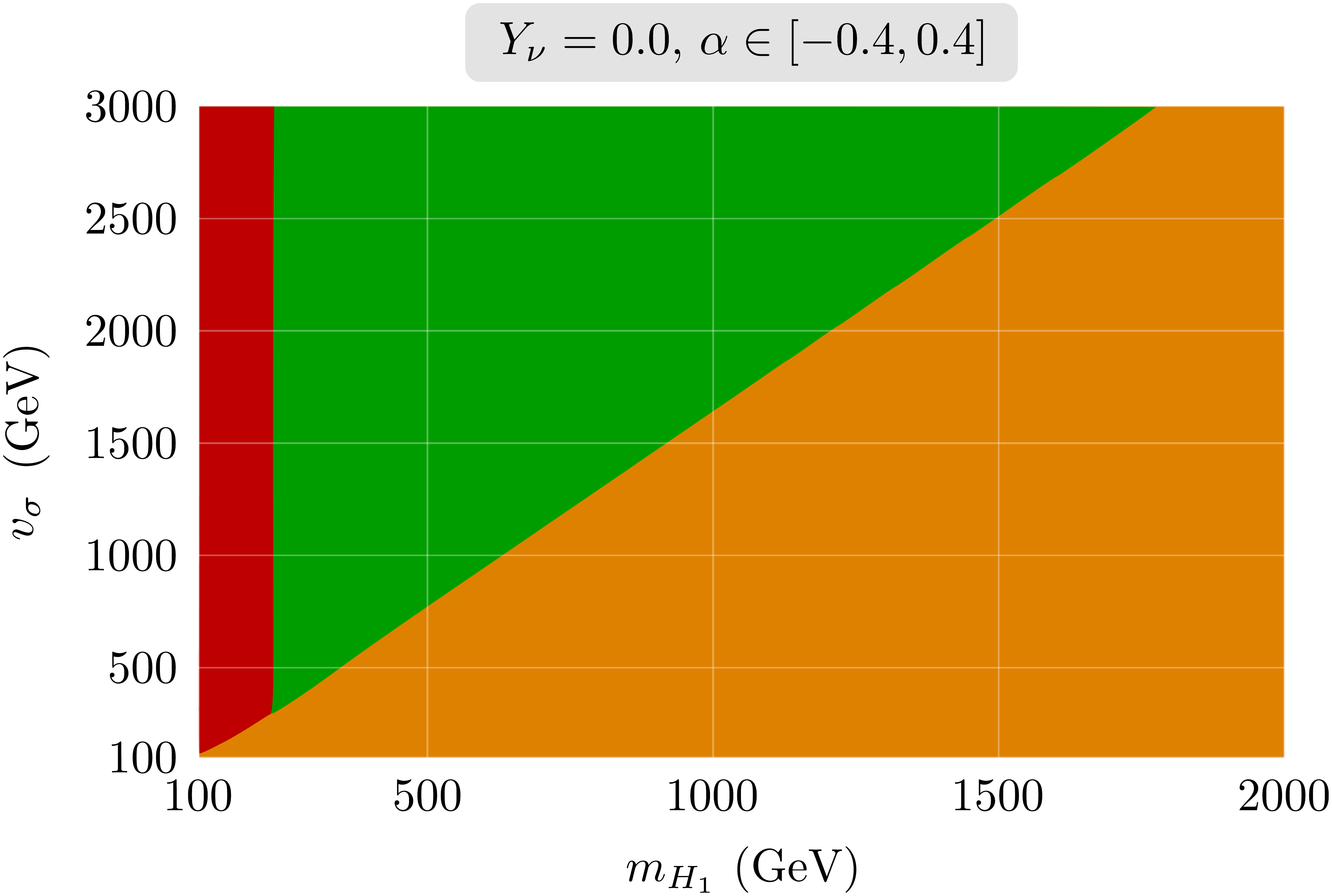}} 
\vspace{-5pt}
\caption{\label{scan1}Values of $m_{H_{2}}$ and $v_{\sigma}$ leading
  to a potential bounded from below (in green on top), a Landau pole
  at some energy scale (in orange, next), or an unstable potential (in
  red, last).}
\end{figure}
 \begin{figure}[htb]
\vspace{-10pt}
  \centering
  {\includegraphics[width=0.48\textwidth]{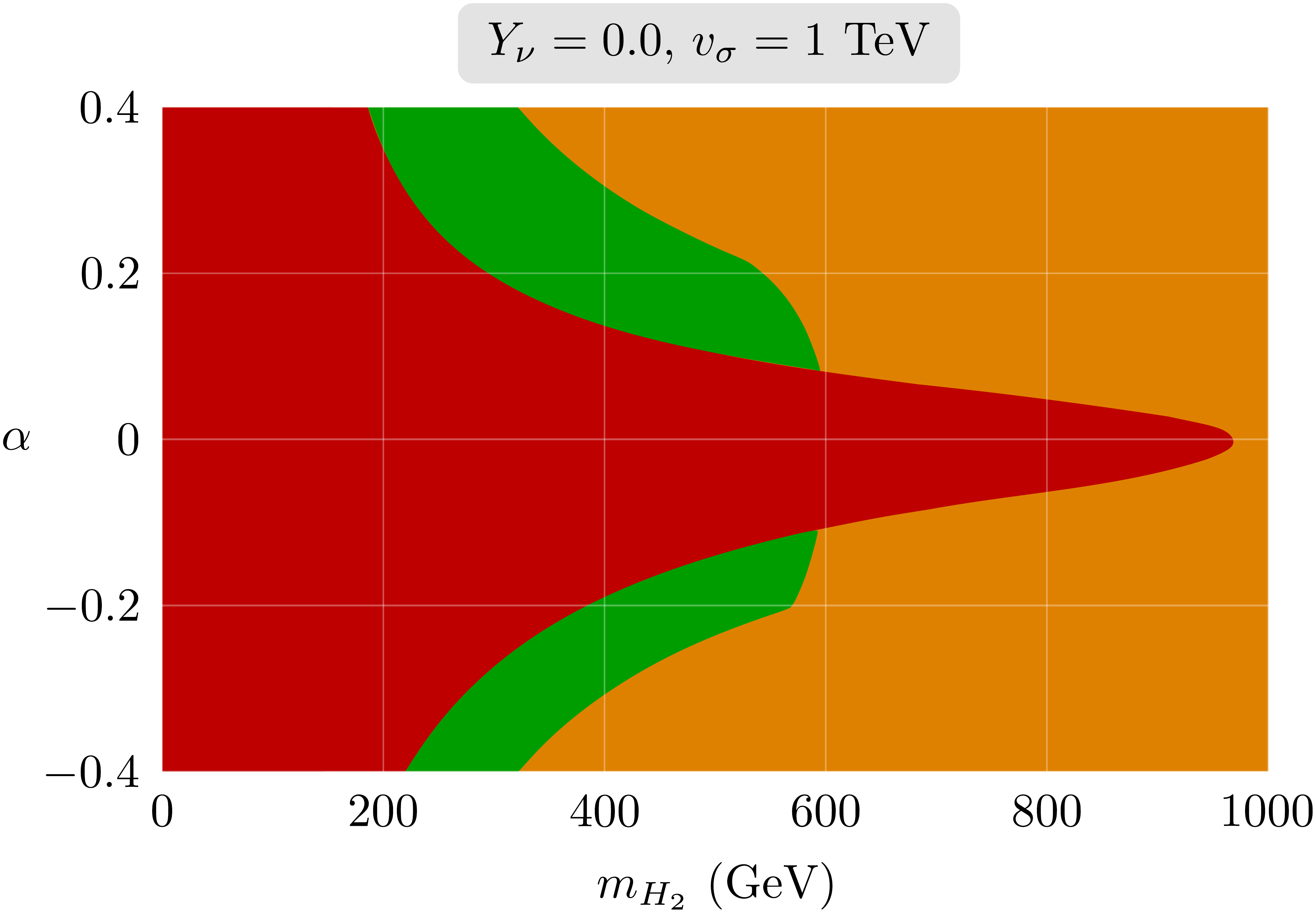}}   \vspace{5pt} \quad
  {\includegraphics[width=0.48\textwidth]{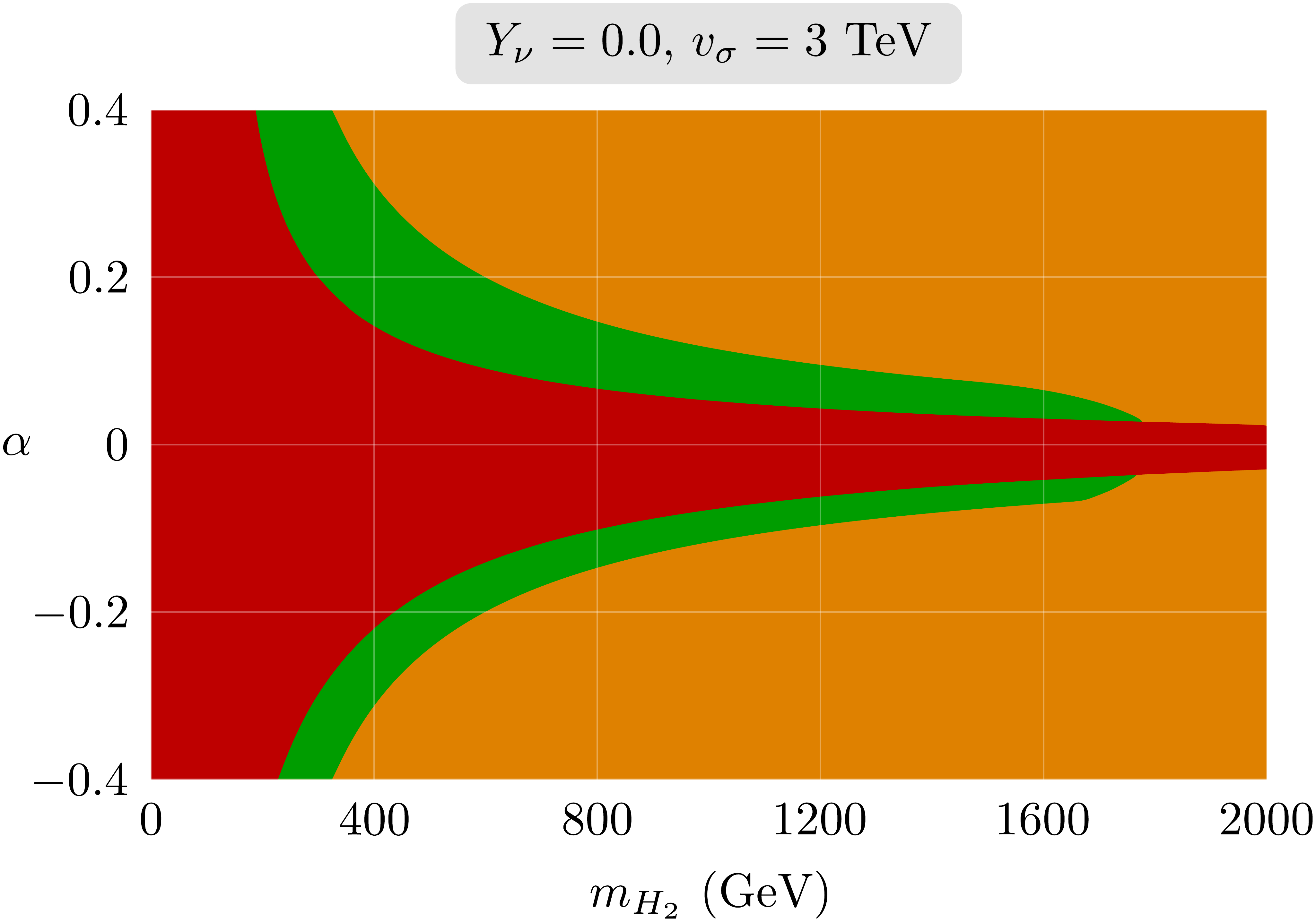}} \\
\vspace{-5pt}
\caption{\label{scan2}Values of $m_{H_{2}}$ and $\alpha$ leading to a
  potential bounded from below (in green), a Landau pole at some
  energy scale (in orange), or an unstable potential (in red).
  Comparing top and bottom panels shows the effect of changing
  $v_{\sigma}$.}
\end{figure}
\section{Neutrino mass generation}

In order to assign to the U(1) symmetry present in Eq.~(\ref{eq:pot}) the
role of lepton number we must couple the new scalar singlet to
leptonic fields. This can be done in a variety of ways. Here we focus
on low-scale generation of neutrino mass~\cite{Boucenna:2014zba}.
For definitiveness we choose to generate neutrino masses through the
inverse seesaw mechanism \cite{Mohapatra:1986bd} with spontaneous \lnv
\cite{gonzalezgarcia:1989rw}.

The fermion content of the Standard Model is augmented by right-handed
neutrinos $\nu_{R}$ (with lepton number +1) and left-handed gauge singlets $S$
(also with lepton number +1) such that the mass term $\nu_{R}^{c}S$ as well as the interactions $SS\sigma$ and  $H\nu_{R}^{c}L$ are
allowed if $\sigma$ carries -2 units of lepton number:\footnote{We ignore for simplicity the extra term
  $\nu_{R}^{c}\nu_{R}^{c}\sigma^{*}$ which is, in principle, also
  allowed.}
\begin{equation}
-\mathscr{L}_{\nu}=Y_{\nu}H\nu_{R}^{c}L+M\nu_{R}^{c}S+Y_{S}SS\sigma+\textrm{h.c.}
\end{equation}
The effective neutrino mass, in the one family approximation, is given
by the expression
\begin{equation}
m_{\nu}=Y_{S}\left\langle \sigma\right\rangle \left(\frac{Y_{\nu}\left\langle H^{0}\right\rangle }{M}\right)^{2}\,,
\end{equation}
which shows that the smallness of the neutrino masses can be
attributed to a small (but natural) $Y_{S}$ coupling, while still
having $Y_{\nu}$ of order one and both $\left\langle
  \sigma\right\rangle $, $M$ in the TeV range.\\

\section{Interplay between neutrino mass and Higgs physics}

\begin{figure}[H]
\vspace{-10pt}
  \centering
  {\includegraphics[width=0.48\textwidth]{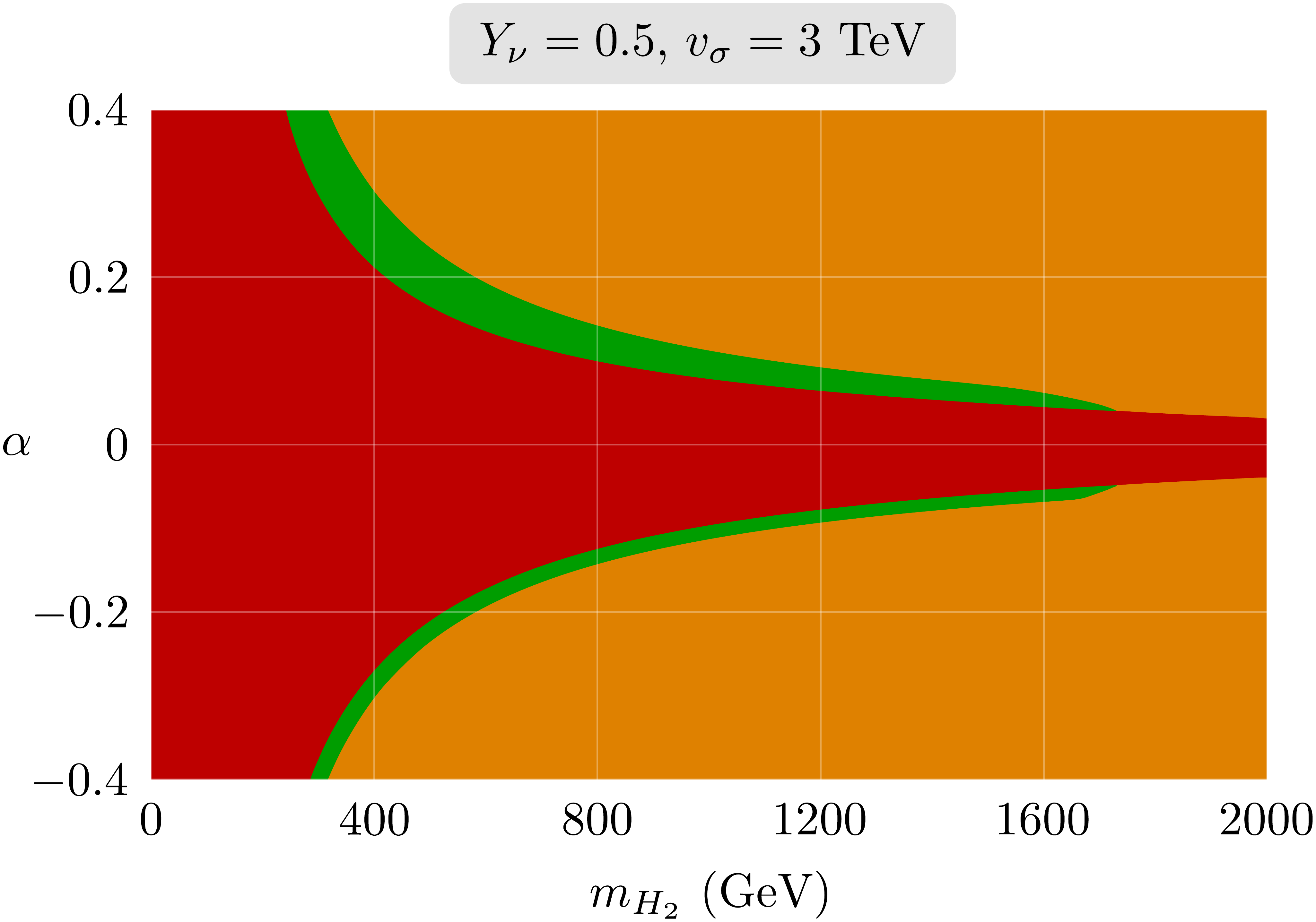}} \\
\vspace{-5pt}
\caption{\label{scan3}Same as in the bottom plot of Fig. \ref{scan2},
  but with $Y_{\nu} \neq 0$.}
\end{figure}

In most cases, the stability of the potential is threatened by the
violation of the condition $\lambda_{2}>0$, as in the Standard Model.
Instability can be avoided with a large $\lambda_{12}$, which might,
however, lead to an unacceptably large mixing angle $\alpha$ between
the two CP-even Higgs mass eigenstates~\cite{Falkowski:2015iwa}. In
such cases, one must rely instead on a heavy $H_{2}$ --- see the green
region in Figs. \ref{scan1}--\ref{scan3}.  Indeed, within the red
regions therein, the potential becomes unbounded from below at some
high energy scale, just like in the Standard Model.  This happens for
relatively small values of either $\alpha$ or $m_{H_{2}}$.
As a result, a tight experimental bound on $\alpha$ can be used to
place a lower limit on the mass of the heavier CP-even scalar.  From
Fig. \ref{scan1} one can also see that the lepton breaking scale
$v_{\sigma}\equiv\sqrt{2}\left\langle \sigma\right\rangle $ must not
be too low, otherwise a big ratio $m_{H_{2}}/\left\langle
  \sigma\right\rangle $ will lead to the existence of a Landau pole in
the running parameters of the model before the Planck scale is reached
(shown in orange).  This also accounts for the difference
between the two plots in Fig.  \ref{scan2}.\\

As far as the neutrino sector is concerned, since $Y_{S}$ is taken to
be small, this parameter has no direct impact on the potential's
stability.  However, it should be noted that in order to obtain
neutrino masses in the correct range, the values of both $v_{\sigma}$
and $Y_{\nu}$ will depend on the one of $Y_{S}$. In principle then,
$Y_{\nu}$ might be large, but not too large, as
$\left|Y_{\nu}\right|\gtrsim0.6$ leads to either unstable or
non-perturbative dynamics. A non-zero Dirac neutrino Yukawa coupling
has a destabilizing effect on the scalar potential which is visible in
the recession of the green region to bigger values of $\alpha$ and
$m_{H_{2}}$, when comparing the bottom plot in Fig.  \ref{scan2} and
the one in Fig. \ref{scan3}.

\begin{figure}[H]
\vspace{-10pt}
  \centering
\includegraphics[width=0.48\textwidth]{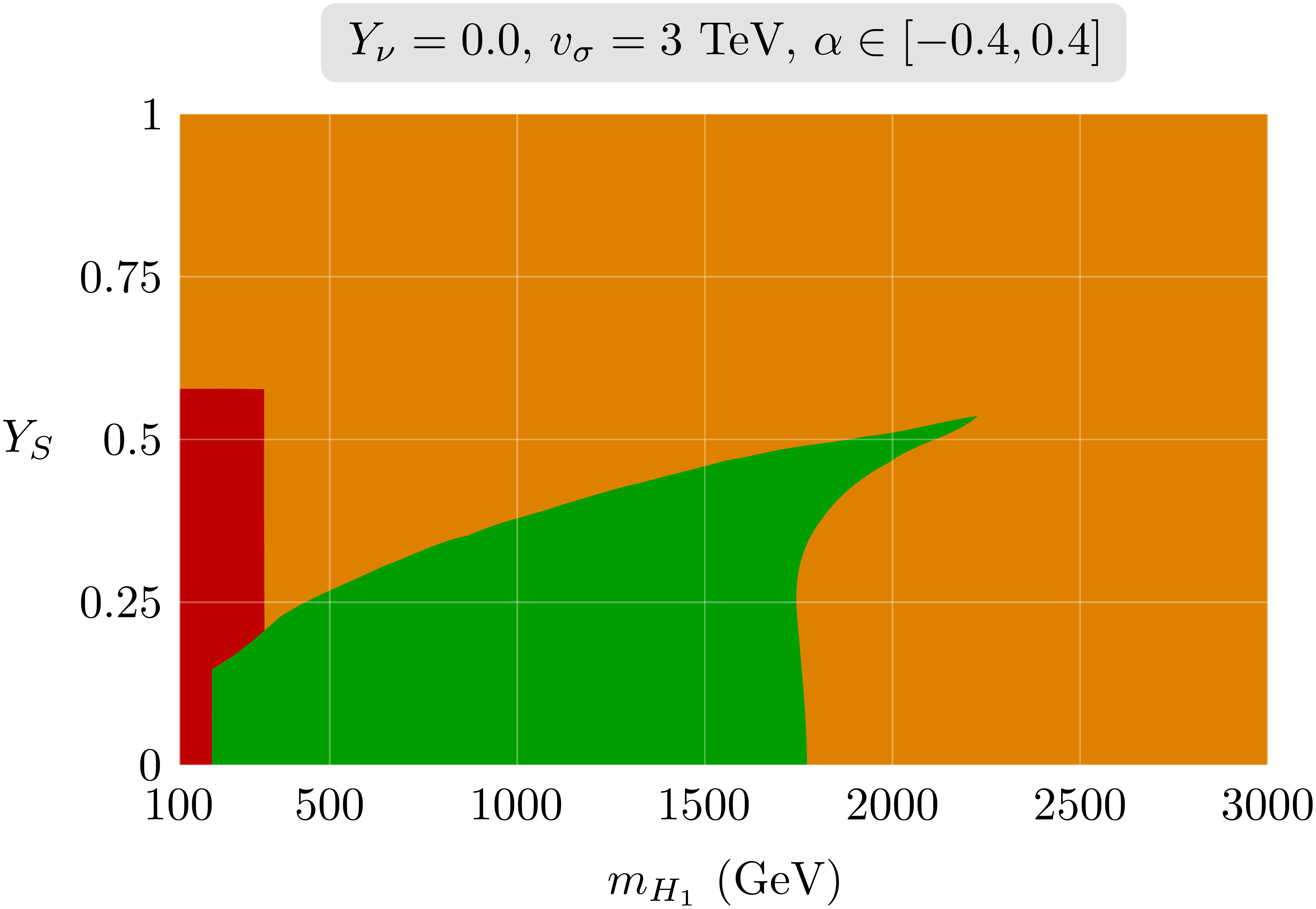}
\vspace{-5pt}
\caption{\label{scan4} The region stability and perturbativity for the
  case of non-zero $Y_S > 0$ and very small $Y_\nu$ is displayed in
  green; the color ordering code is the same as in the scan in
  Fig.~\ref{scan1}.}
\end{figure}

Another interesting possibility is to have a negligible $Y_\nu$ and
potentially sizeable $Y_S$. In this case, if we keep $M$ of the order
of the TeV, we find that the region of stability and perturbativity
(shown in green in Fig. \ref{scan4}) depends significantly upon the
parameter $Y_S$ characterizing spontaneous \lnv and neutrino mass
generation through $\vev{\sigma}$. To be more precise, as shown in
Fig. \ref{scan4} the allowed values for the mass of the heavy scalar
boson ($m_{H_2}$) varies with this Yukawa coupling; for example, if
$m_{H_2}$ was to be found to be, say, $\sim 2$ TeV ($v_\sigma =3$ TeV
by assumption here), then one would conclude that either $Y_S\sim0.5$
or the scalar sector must be strongly interacting.\\

\section{Conclusions}

In conclusion, the Standard Model vacuum is unstable for the measured
top and Higgs boson masses.
However the theory is incomplete as it has no masses for neutrinos. We
have therefore generalized its symmetry breaking potential in order to
induce naturally small neutrino masses from the breaking of lepton
number.
We have examined the vacuum stability issue in schemes with
spontaneous breaking of global lepton number at the TeV scale, showing
how one can naturally obtain a consistent behavior of the scalar
potential at all scales, avoiding the vacuum instability.
Given that the new physics parameters of the theory are not known, it
sufficed for us to adopt one-loop  renormalization group equations.
Since all new particles in the model lie at the TeV scale, they can be
probed with current experiments, such as the LHC. Invisible decays
of the two CP-even Higgs bosons, $H_i\to JJ$, were discussed
in~\cite{Bonilla:2015uwa}. Improved sensitivity is expected from the
13~TeV run of the LHC. In addition, we expect enhanced rates for
lepton flavour violating
processes~\cite{bernabeu:1987gr,Deppisch:2004fa,Deppisch:2005zm}.
In summary, schemes such as the one explored in this letter may shed
light on two important drawbacks of the \sm namely, the instability
associated to its gauge symmetry breaking mechanism and the lack of
neutrino mass.\\ 

\begin{center}
{\bf Acknowledgements}\\
\end{center}

We thank Sofiane Boucenna for discussions in the early phase of this
project. This work was supported by MINECO grants FPA2014-58183-P,
Multidark CSD2009-00064 and the PROMETEOII/2014/084 grant from
Generalitat Valenciana.\\[-.2cm]

\section{Appendix A}
In this appendix we provide some details on the scalar sector of the
model. The potential in equation (\ref{eq:pot}) is controlled by 5
parameters ($\mu_{1}^{2}$, $\mu_{2}^{2}$, $\lambda_{1}$,
$\lambda_{2}$, and $\lambda_{12}$) which one can translate into two
vacuum expectation values ($v_{\sigma}=\sqrt{2}\left\langle
  \textrm{Re}\left(\sigma\right)\right\rangle $ and
$v_{H}=\sqrt{2}\left\langle \textrm{Re}\left(H^{0}\right)\right\rangle
$), two mass eigenvalues ($m_{H_{1}}$ and $m_{H_{2}}$) and a mixing
angle $\alpha$:
\begin{align}
	\lambda_{1} & =\frac{m_{H_{1}}^{2}\cos^{2}\alpha+m_{H_{2}}^{2}\sin^{2}\alpha}{2v_{\sigma}^{2}}\,,\label{eq:Lsigma}\\
	\lambda_{2} & =\frac{m_{H_{1}}^{2}\sin^{2}\alpha+m_{H_{2}}^{2}\cos^{2}\alpha}{2v_{H}^{2}}\,,\\
	\lambda_{12} & =\frac{\left(m_{H_{1}}^{2}-m_{H_{2}}^{2}\right)\cos\alpha\sin\alpha}{v_{\sigma}v_{H}}\,,\\
	-\mu_{1}^{2} & =\frac{v_{H}\cos\alpha\sin\alpha\left(m_{H_{1}}^{2}-m_{H_{2}}^{2}\right)+m_{H_{1}}^{2}v_{\sigma}\cos^{2}\alpha+m_{H_{2}}^{2}v_{\sigma}\sin^{2}\alpha}{2v_{\sigma}}\,,\\
	-\mu_{2}^{2} & =\frac{v_{\sigma}\cos\alpha\sin\alpha\left(m_{H_{1}}^{2}-m_{H_{2}}^{2}\right)+m_{H_{1}}^{2}v_{H}\sin^{2}\alpha+m_{H_{2}}^{2}v_{H}\cos^{2}\alpha}{2v_{H}}\,,\label{eq:mu2H}
\end{align}
with
\begin{align}
	\left(\begin{array}{c}
		H_{1}\\
		H_{2}
	\end{array}\right) & \equiv\left(\begin{array}{cc}
	\cos\alpha & \sin\alpha\\
	-\sin\alpha & \cos\alpha
\end{array}\right)\left(\begin{array}{c}
\sqrt{2}\textrm{Re}\left(\sigma\right)\\
\sqrt{2}\textrm{Re}\left(H^{0}\right)
\end{array}\right)\,.
\end{align}
On the other hand, it is well known that the Standard Model potential
is controlled by just two parameters $\mu^{2}$ and $\lambda$:
\begin{align}
	V_{SM}\left(H\right) & =\mu^{2}\left(H^{\dagger}H\right)+\lambda\left(H^{\dagger}H\right)^{2}\,.
\end{align}
For a reasonably small mixing angle $\alpha$, one can consider that
the state $H_{1}$ is mostly made of the real part of the singlet,
hence we may integrate out $\sqrt{2}\textrm{Re}\left(\sigma\right)$. In this
approximation, we note that 
\begin{align}
	\lambda & \approx\lambda_{2}-\frac{\lambda_{12}^{2}}{4\lambda_{1}}\,,\\
	\mu^{2} & \approx\mu_{2}^{2}-\frac{\lambda_{12}}{2\ensuremath{\lambda_{1}}}\mu_{1}^{2}\,,
\end{align}
at the scale of decoupling, meaning in particular that there is a
tree-level threshold correction between $\lambda_{2}$ and the Standard
Model quartic coupling $\lambda$. For the results in this paper, we
neglect altogether the small Standard Model range between the $m_{Z}$
and $m_{H_{1}}$ scale, starting instead with equations
(\ref{eq:Lsigma})--(\ref{eq:mu2H}), which already include this
threshold effect.  
\section{Appendix B}

For completeness, we write down here the renormalization group
equations of the model parameters which are relevant for the study of
the potential's stability. We work with the 1-family approximation,
ignoring the bottom and tau Yukawa couplings. These equations were
obtained with the SARAH program \cite{Staub:2015kfa} (see also
\cite{Lyonnet:2013dna}) and explicitly checked by us using the results
in \cite{Luo:2002ti}; furthermore they are consistent with
\cite{EliasMiro:2012ay}. As usual, $t$ stands for the natural
logarithm of the energy scale.

\vspace{-3pt}
\thinmuskip=1mu
\medmuskip=1.5mu 
\thickmuskip=2.0mu
\begin{eqnarray}
\left(4\pi\right)^{2}\frac{dg_{i}}{dt} & = &b_{i}g_{i}^{3} \text{ with}\ \ b_{i}=\left(\frac{41}{10},-\frac{19}{6},-7\right)\,,\\
\left(4\pi\right)^{2}\frac{dY_{t}}{dt} & = &\left(-\frac{17}{20}g_{1}^{2}
-\frac{9}{4}g_{2}^{2}-8g_{3}^{2}+\frac{9}{2}Y_{t}^{2}+Y_{\nu}^{2}\right) Y_{t}\,,\\
\left(4\pi\right)^{2}\frac{dY_{\nu}}{dt} & = &\left(-\frac{9}{20}g_{1}^{2}
-\frac{9}{4}g_{2}^{2}+3Y_{t}^{2}+\frac{5}{2}Y_{\nu}^{2}\right)  Y_{\nu} \,,\\
\left(4\pi\right)^{2}\frac{dY_{S}}{dt} & = &6Y_{S}^{3}\,,\\
\left(4\pi\right)^{2}\frac{d\lambda_{1}}{dt}  & = &20\lambda_{1}^{2}+
2\lambda_{12}^{2}+8\lambda_{1}Y_{S}^{2}-16Y_{S}^{4}\,,\\
\left(4\pi\right)^{2}\frac{d\lambda_{2}}{dt} & = &\frac{27}{200}g_{1}^{4}
+\frac{9}{20}g_{1}^{2}g_{2}^{2}+\frac{9}{8}g_{2}^{4}\notag\\
& - &\left(\frac{9}{5}g_{1}^{2}+9g_{2}^{2}\right)\lambda_{2}+
24\lambda_{2}^{2}+\lambda_{12}^{2} \\
& + &\lambda_{2}\left(12Y_{t}^{2}+4Y_{\nu}^{2}\right)-\left(6Y_{t}^{4}
+2Y_{\nu}^{4}\right)\,,\notag\\
\left(4\pi\right)^{2}\frac{d\lambda_{12}}{dt} & = &\left[-
\left(\frac{9}{10}g_{1}^{2}+\frac{9}{2}g_{2}^{2}\right)+6Y_{t}^{2}
+2Y_{\nu}^{2}\right.\notag\\
& + &\left.4Y_{S}^{2}+8\lambda_{1}+12\lambda_{2}+4\lambda_{12} \vphantom{\left(\frac{9}{10}g_{1}^{2}+\frac{9}{2}g_{2}^{2}\right)}\right]
\lambda_{12}\,.
\end{eqnarray}

\end{document}